\documentclass[cits]{PoS}

\title{Small and young radio sources}

\ShortTitle{Small and young radio sources}

\author{\speaker{Monica Orienti}\thanks{This work has benefited from 
research funding from the European
Community's sixth Framework Programme under RadioNet R113CT 2003 5058187}\\
        IAC - c/ Via Lactea s/n, E-38205 La Laguna (Tenerife), Spain\\
        E-mail: \email{orienti@ira.inaf.it}}

\abstract{It is currently accepted that compact and bright radio
  sources characterized by a convex spectrum peaking at frequencies
  ranging from 100 MHz to a few GHz are young objects. In this
  scenario, high frequency peaker (HFP) radio sources, with a turnover
  frequency higher than 5 GHz are good candidates to be extremely young
  radio sources with ages of up to a few thousand years. The knowledge 
  of the conditions in young radio source is fundamental in
  order to draw reliable evolution models able to describe the entire
  life-cycle of the radio emission.  
  Given the high spatial
  resolution and the large frequency range spanned, VLBI observations
  provide a unique opportunity to constrain the physical conditions in
  young radio sources, and to investigate the role played by the
  environment on the source growth. 
Recent VLBI results are
  discussed here.}

\FullConference{The 9th European VLBI Network Symposium on The role of VLBI in the Golden Age for Radio Astronomy and EVN Users Meeting\\
		 September 23-26, 2008\\
		 Bologna, Italy}

\begin{document}

\section{Small and young radio sources}

An active galactic nucleus (AGN) is a compact and luminous 
region hosted in the centre of a galaxy. Its luminosity, much higher
than that observed in ``non-active'' galaxies, is not due to stellar
emission, but it is related to accretion mechanisms on the central
black hole. Only a small fraction of the AGN population ($\sim$ 10\%)
is radio-loud, making the origin of the radio emission
a still open question.\\
The onset of radio emission is currently considered related to mergers
or accretion events taking place in the host galaxy. In the
early-stages of the radio source evolution, the radio emission has to
make its way through the dense interstellar medium (ISM) before plunging
into the smoother intracluster medium (ICM), and giving origin to extended (up
to a few Mpc) radio sources.
In this scenario, the linear size of the radio
source gives an indication of the evolution stage (i.e. the age) of the
radio emission.\\
A significant fraction (15-30\%) of sources in
flux-density limited radio catalogues is represented by powerful
(P$_{\rm 1.4 GHz} > 10^{25}$ W/Hz) and intrinsically 
compact ($<$ 1$^{\prime\prime}$ ) radio
sources, with a convex synchrotron radio spectrum peaking at
frequencies between $\sim$100 MHz and a few GHz. 
If the peak of the spectrum occurs either at a few hundreds MHz or around one
GHz, the radio sources are identified as ``compact steep spectrum
(CSS) or ``GHz-peaked spectrum'' (GPS) respectively. \\
When imaged with high
spatial resolution, they usually display a two-sided morphology
dominated by hot-spots and mini-lobes. For this reason, these radio
sources are known as ``Compact Symmetric Objects'' (CSOs).\\
Their small size, together with 
their morphology, a scaled-down version (0.1 - 15 kpc) of the extended
classical radio galaxies, suggest
that these objects represent a young stage in the individual radio
source evolution (see e.g. \cite{fanti95, snellen00}). This interpretation is supported by the detection of
hot spot advance speed in a dozen of the most compact CSOs, leading to
a kinematic age of about $10^{3}$ years \cite{pc03}. The young age of
CSOs has a further confirmation by means of spectral studies 
\cite{murgia99, murgia03} providing radiative ages of
$10^{3} - 10^{5}$ years.\\
The competing model, known as the {\it frustration scenario}, postulates
that these objects are small because their are trapped by an
extremely dense ambient medium able to frustrate the source
growth \cite{vbreugel84}. 
However, no observational evidence of the presence of an
unusually dense environment has been found \cite{fanti00, aneta05},
giving an indirect support to the {\it youth scenario}.\\
The main property characterizing small and young radio sources is
the presence of a peak in their synchrotron spectrum. 
This feature is commonly
explained by means of synchrotron-self absorption (SSA)
\cite{snellen00}. The empirical anti-correlation found between the linear
size and the peak frequency \cite{odea97} supports this view: as
the source grows, its radio emission becomes optically-thin at lower and lower
frequencies. This implies that the smallest, and thus youngest radio
sources, must be sought among those with a spectrum peaking at high
frequency, above a few GHz. With their spectral peak occurring well
above 4-5 GHz, high frequency peakers (HFP) are the best candidate to
be newly born radio sources \cite{dd00, dd03}.\\
Since the selection of young radio sources is based mainly on the
position of the spectral peak, it is possible that different kind of
objects, like blazars may contaminate the samples. In this objects, the
compact size is due to projection effects which
foreshorten their real linear size, and it is not related to the source
age. To discriminate between young and boosted objects, an accurate
analysis of the radio properties is unavoidable.\\

\section{Radio properties}

\subsection{Spectral variability}

The presence of a convex spectrum with a peak occurring at different
frequencies depending on the age of the source, is the main
characteristic of young radio sources. In the presence of SSA, the
peak is expected to move to lower frequency and the flux density in
the optically-thick part of the spectrum should increase
as the source grows in size,
without indication of significant variability in the spectrum on small
temporary scale. Multi-epoch observations showed that CSS, GPS
and HFP sources identified with galaxies, are the least variable class of
extragalactic objects \cite{odea98, stanghella98, orienti07}, 
while the majority of those
identified with quasars show flux-density and spectral-shape variability
\cite{torniainen05, tinti05}. In a
significant fraction of GPS/HFP quasars, 
the radio spectrum
turns out to be flat in a subsequent observing epoch
\cite{torniainen05, orienti07},
indicating a different nature for this class of objects. \\

\subsection{Morphology}

\begin{figure}
\begin{center}
\includegraphics{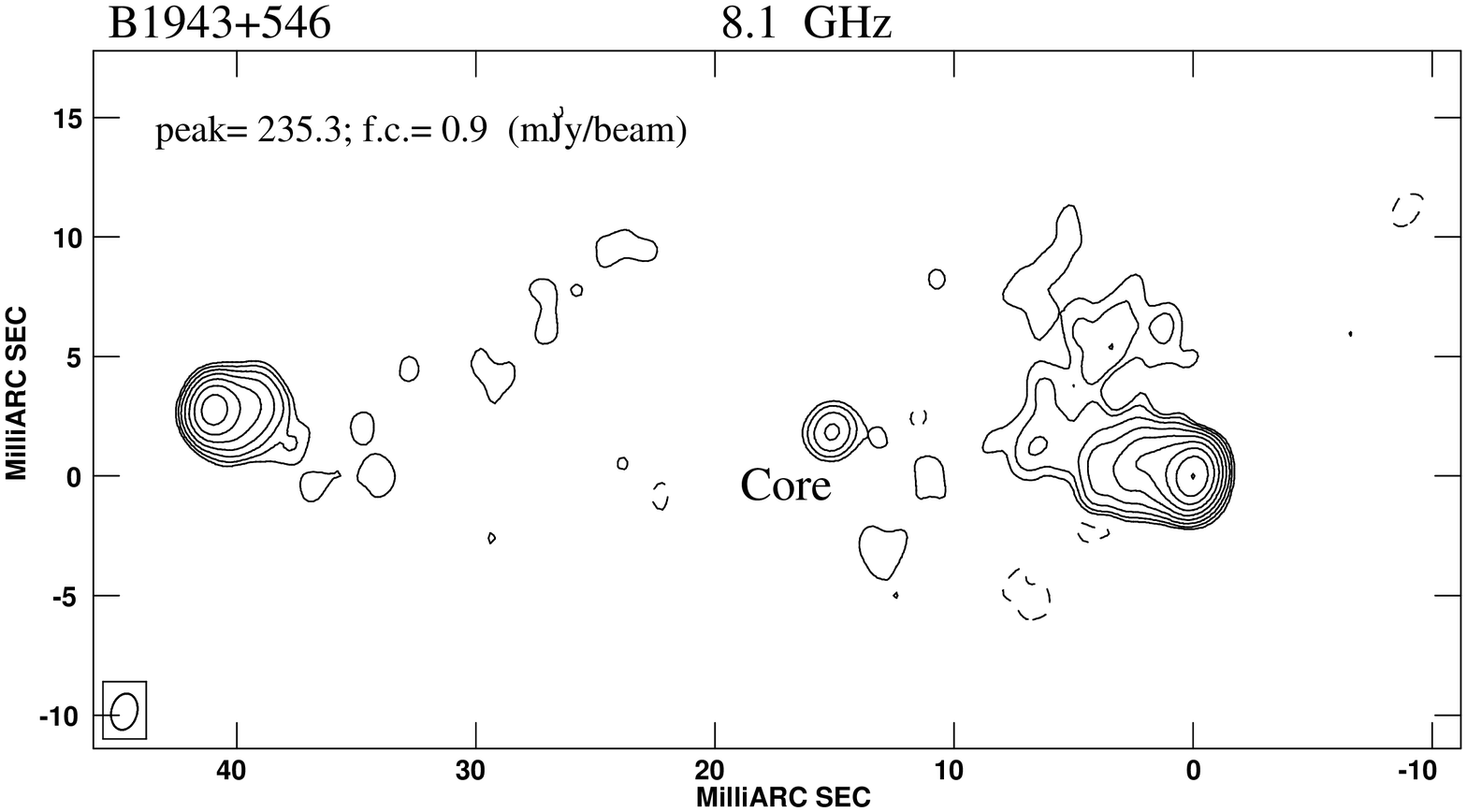}
\includegraphics{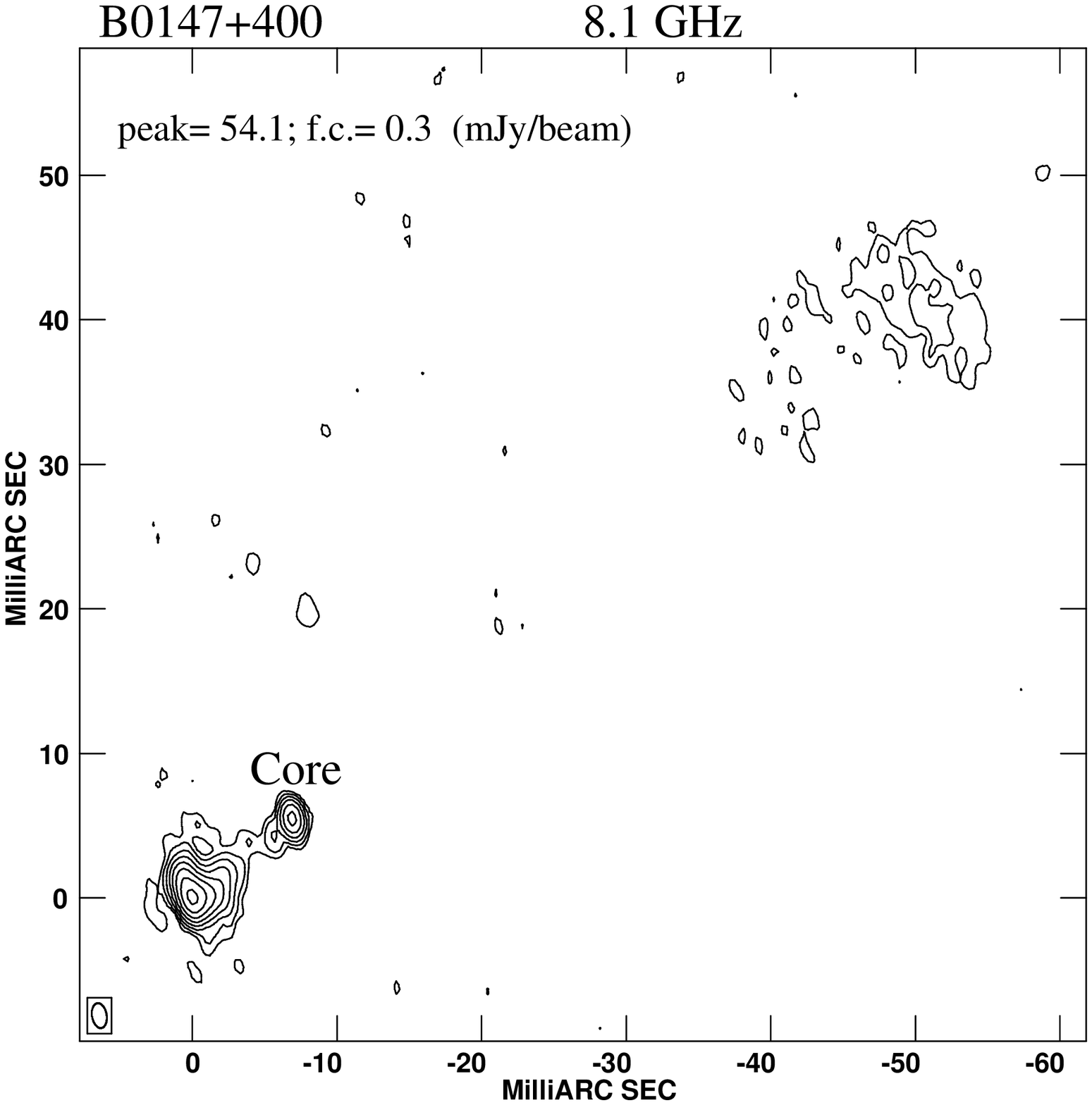}
\vspace{5cm}
\caption{Example of ``symmetric'' structures in CSS/GPS radio sources.}
\label{symm}
\end{center}
\end{figure}

Given the sub-arcsecond size of young radio sources, the study of
their structure became possible with the advent of VLBI. The first
pioneering VLBI works on CSS/GPS sources revealed simple symmetric
structures. Deeper observations allowed a more accurate description of
their morphology \cite{fanti90}. 
Young radio sources, at least those identified with
galaxies, have a ``symmetric'' structure dominated by two well-separated
components with steep spectral index, interpreted as mini-lobes and/or
hot spots, occasionally with a weak
unresolved component in between, accounting for a small fraction
($<$1\%) of the flux density, usually detected at high frequency
only, which may contain the source core (Fig. \ref{symm}a). 
In this context, the term ``symmetric'' only refers to a two-sided
morphology  and does not imply that the source components have similar
characteristics. Between the ``symmetric'' objects there are radio sources
with highly asymmetric structures (Fig. \ref{symm}b). In a significant
fraction ($\sim$65\%, \cite{saikia03}) of small radio galaxies, the
two lobes have very different luminosity. Furthermore, when the
source core is detected, it has been observed that also the separation
ratio of the outer components from the nucleus is asymmetric, being
the brightest also the closest to the core. Such
asymmetries are more commonly found in small radio sources, than 
in large radio galaxies \cite{longair00}, suggesting an influence
played by the ambient medium on the source growth.\\ 
On the other hand, small radio
sources identified with quasars mainly display a one-sided Core-Jet
morphology with a straight or bent structure always with a well
visible core accounting for the majority of the flux
density, and located at one edge of the source. The Core-Jet
morphology is the structure usually displayed by beamed blazar
objects. Small radio sources with such a morphology also
show significant flux density and spectral-shape variability
\cite{orienti07}, suggesting, again, a different interpretation for the 
radio emission.\\ 
  
\subsection{Polarization}

The study of the polarization enables us to derive the distribution of
the magnetic field in young radio sources. However, 
young radio sources completely reside within the ISM
of the host galaxy. This gas can act as a Faraday Screen
causing substantial depolarization of the synchrotron radiation.\\
In large ($>$ 5 kpc) CSS objects, the radiation has high percentage of
polarization ($>$10\%, \cite{fanti04}) and the magnetic field is
generally parallel to the jet axis \cite{fanti01}, as found in
edge-brightened large radio galaxies. On the other hand, 
radio galaxies smaller than 5 kpc are
unpolarized or possess very low percentage of polarized
emission \cite{fanti04, orienti08a}. 
The high rotation measures (RM) found suggests that the
radiation is strongly depolarized by the magneto-ionic plasma of the
ambient medium. \\
As found for variability and morphological properties, 
a different behaviour is shown by GPS and HFP quasars, where high
values of polarized emission are detected \cite{orienti08a,
  stanghella98}. \\

The different characteristics shown by small radio sources with
different optical identification suggest that  
small radio objects hosted either by 
galaxies or by quasars represent two different radio source populations:
genuinely young radio objects the former, beamed blazars the
latter.\\   

\section{Physical conditions}

The radio emission
of extragalactic objects is due to synchrotron radiation produced by
relativistic electrons with a power-law energy distribution. A curvature
in the spectrum produced by  
synchrotron self-absorption mechanism is thus present by default, and
the frequency at which it occurs is strongly related to the size of the
radio component. In the context of evolution models, this explain
the anti-correlation found between peak frequency and linear size:
as the source expands, the synchrotron radiation becomes optically
thin at lower energy, shifting the radio spectrum towards lower
frequency. \\
In order to draw a complete picture able to describe the various stages
of the radio source evolution, a deep knowledge of the physical
conditions in young radio sources is necessary.
Evolution models proposed so far, e.g.  \cite{fanti95, snellen00}, are
based on the assumption of minimum energy conditions, corresponding to
a near equipartition of energy between particles and magnetic field
\cite{pacho70}. Equipartition magnetic fields derived in young radio
sources ranges from a few mG, in the compact components of CSS
\cite{fanti95}, up to
tens or hundreds mG in HFPs \cite{orienti08b}. 
However, there is no a-priori reason for believing that 
magnetic fields are those of equipartition, and it is important to
derive the magnetic field in an independent way.\\
If the spectral peak is due to SSA as stated before, the magnetic
field $B$ of a homogeneous synchrotron component 
can be computed by observable quantities only:

\begin{equation}
B = f( \alpha )^{-5} \theta^{4} \nu_{\rm max}^{5} S_{\rm max}^{-2}
(1+z)^{-1}
\label{ssa} 
\end{equation}

\noindent where $\theta$ is the source angular size, $S_{\rm max}$ is
the peak flux density, $\nu_{\rm max}$ the peak frequency, $z$ is the
redshift and $f(\alpha$) is a function of the spectral index
\cite{kellerman81}. \\ 
The main difficulty in applying this method is the uncertainty in
determining the source parameters, in particular the peak frequency
and the angular
size. For an accurate determination of the peak frequency, the radio
spectrum must be well sampled in both its optically-thick and -thin
parts. Furthermore, the magnetic field should not be computed on the
whole source, but on individual sub-structures, which are a better
approximation of homogeneous components.\\ 
The VLBI has the ideal frequency coverage and spatial resolution to
perform such an investigation.\\
Magnetic fields derived from the spectral peak of CSS radio sources
were within a factor of 16 of the equipartition values
\cite{scott77}. The discrepancy between the values may arise from the
uncertainties in the determination of the spectral peak which occurs at
rather low frequencies where the sampling is more difficult and size
determination more critical.\\
In HFP objects, the spectral peak at a few GHz allows an accurate
determination of the peak parameters. Furthermore, given their compact
size due to the severe radiative losses which prevent the formation of
extended lobe-like features, 
HFP sub-structures 
can be considered a good approximation of uniform synchrotron components. 
The magnetic fields inferred are
in good agreement with those computed assuming equipartition, with a
few exceptions \cite{orienti08b}. 
In some HFP components, the magnetic fields derived are above a few
Gauss, that is one or two orders of magnitude higher than the equipartition
value. However, in these components it was noted that 
the optically-thick part of the spectrum has a
spectral index $\alpha < -2.5$ ($S \propto \nu^{- \alpha}$), i.e. the
canonical value of SSA, implying the presence of an additional
contribution from free-free absorption (FFA). In this case, the
magnetic fields derived by means of the peak parameters are physically
meaningless. \\  
%

\section{Ambient medium}

The origin of radio activity in AGN is thought to be related to the
availability of fuel to feed the central engine, likely provided by
merger events. For this reason, galaxies hosting a radio source in an early
evolutionary stage, are supposed to possess a rich and dense
interstellar medium. The presence of significant amount of gas in young
radio sources was proved by the high incidence of H{\small I} absorption
\cite{ylva03}, which is substantially higher than what typically found
in large (hundreds of kpc in size) radio galaxies \cite{morganti01}.\\
Statistical studies of the properties of the H{\small I} absorption in young
CSS/GPS objects pinpointed the existence of an anti-correlation
between the H{\small I} column density and the linear size \cite{ylva03,
  gupta06}: the smaller the source, the larger the H{\small I} column
density. This result can be explained assuming that the neutral
hydrogen is organized in a circumnuclear structure, such as a torus or
a disk. In this context, the H{\small I} absorption is detected against the
receading jet, when our line of sight passes through the circumnuclear
disk/torus along its way towards the radio emission
\cite{mundell03}. As the source grows, the background emission of
the counter-jet moves toward the outer and less dense part of the
atomic structure, causing a decrement of the H{\small I} column
density. \\
Surprisingly, 
the anti-correlation found in CSS/GPS objects is not present in the
small HFP sources \cite{orienti06}.
This unexpected result can be explained again in a torus
scenario. Given the extremely small size of HFPs, the line of sight
to the counter-jet passes mainly through the ionized part of the ISM,
piercing the atomic structure in its innermost part, where it is
hotter and less dense. Support to this interpretation arises from the
detection of large amount gas in HFPs \cite{dallacasa09,
  orienti08b}, responsible for free-free absorption.\\
The anti-correlation between linear size and H{\small I}
column density was derived by observations with a spatial
resolution not adequate to determine the distribution of the atomic gas.
To test whether the neutral hydrogen is settled in a circumnuclear
structure, the high spatial resolution provided by VLBI observations
are needed.\\
In the GPS sources 3C\,31.04 \cite{conway96} and 1946+708 \cite{peck99} a
torus-like structure has been detected. On the other hand, in the
GPS sources 4C\,12.50 \cite{morganti04} and 0402+379
\cite{maness04}, and in the two very asymmetric 
CSS 3C\,49 and 3C\,268.3 \cite{labiano06},
VLBI observations showed that H{\small I} absorption occurs far away
from the centre, indicating that the absorption is likely due to an
off-nuclear cloud. Furthermore, in 3C\,49 and 3C\,268.3, H{\small I}
absorption is detected only against the brightest lobe, which turned
out to be also the closest to the core. This suggests that 
jet-cloud
interactions may take place, at least when the radio emission is still
growing within the ISM of the host galaxies. Such an impact
may slow down the propagation of the interacting jet and enhance its
luminosity, explaining in this way the strong luminosity an arm-length
asymmetries observed in many young radio sources.  

\section{Conclusions}
Flux density limited catalogues are known to possess a significant
fraction of small radio sources showing a peak in their synchrotron
radio spectra. Their small size, together with their two-sided
morphology, a scaled-down version of the structures shown by large
radio galaxies, suggest that these radio sources represent an early
stage in the evolution of the radio emission. 
The study of their radio properties, namely the spectral variability,
morphology and polarization, pointed out that those objects optically
identified with galaxies are genuinely young radio sources, while the
majority of those identified with quasars are likely boosted blazar
objects.\\
The presence of a spectral peak is the main characteristic of young
radio sources. This peak is due to SSA, although an additional
contribution from FFA has been found in the most compact objects.
Where the spectral peak is originated by SSA only, it is possible
to derive the magnetic field by means of the peak parameters. Magnetic
fields computed in this way are generally in good
agreement with those calculated assuming
equipartition of energy between particles and magnetic field, 
providing a strong evidence that young radio
sources are in minimum energy conditions. \\
The presence of a dense and inhomogeneous ambient medium surrounding
young radio sources has been pinpointed by frequent detection of
neutral hydrogen in absorption. Although in many cases the atomic gas
is organized in settled, circumnuclear structures as disks and tori,
VLBI observations showed the presence of unsettled, off-nuclear atomic
clouds, responsible for H{\small I} absorption in the outer regions of
the radio sources. The interaction between one of the jet and an
off-nuclear cloud may influence the source growth, for example slowing
down the source expansion, and producing asymmetric structures. \\
Future telescopes with good sensitivity and resolution like SKA, will
provide us with new information on the source environment, that is a
fundamental ingredient for the definition of a complete picture of the
radio source evolution.\\

\end{document}